# Deterministic creation and deletion of a single magnetic skyrmion observed by direct time-resolved X-ray microscopy

Seonghoon Woo,[1†*] Kyung Mee Song,[1,2†] Xichao Zhang,[3†] Motohiko Ezawa,[4] Yan Zhou,[3] Xiaoxi Liu,[5] Markus Weigand,[6] S. Finizio,[7] J. Raabe,[7] Min-Chul Park,[8] Ki-Young Lee,[1] Jun Woo Choi,[1,9] Byoung-Chul Min,[1,9] Hyun Cheol Koo,[1,10] and Joonyeon Chang[1,9]

[1]Center for Spintronics, Korea Institute of Science and Technology, Seoul 02792, Korea
[2]Department of Physics, Sookmyung Women's University, Seoul 04130, Korea
[3]School of Science and Engineering, The Chinese University of Hong Kong, Shenzhen, Guandong 518172, China
[4]Department of Applied Physics, The University of Tokyo, 7-3-1 Hongo, Tokyo 113-8656, Japan
[5]Department of Electrical and Computer Engineering, Shinshu University, 4-17-1 Wakasato, Nagano 380-8553, Japan
[6]Max-Planck-Institut für Intelligente Systeme, 70569 Stuttgart, Germany
[7]Swiss Light Source, Paul Scherrer Institut, 5232 Villigen, Switzerland
[8]Center for Opto-Electronic Materials and Devices, Korea Institute of Science and Technology, Seoul 02792, Korea
[9]Department of Nanomaterials Science and Engineering, Korea University of Science and Technology, Daejeon 34113, Korea
[10]KU-KIST Graduate School of Converging Science and Technology, Korea University, Seoul 02481, Korea

† These authors contributed equally to this work.
* Authors to whom correspondence should be addressed: shwoo_@kist.re.kr

**Spintronic devices based on magnetic skyrmions are a promising candidate for next-generation memory applications due to their nanometre-size, topologically-protected stability and efficient current-driven dynamics. Since the recent discovery of room-temperature magnetic skyrmions, there have been reports of current-driven skyrmion displacement on magnetic tracks and demonstrations of current pulse-driven skyrmion generation. However, the controlled annihilation of a single skyrmion at room temperature has remained elusive. Here we demonstrate the deterministic writing and deleting of single isolated skyrmions at room temperature in ferrimagnetic GdFeCo films with a device-compatible stripline geometry. The process is driven by the application of current pulses, which induce spin-orbit torques, and is directly observed using a time resolved nanoscale X-ray imaging technique. We provide a current-pulse profile for the efficient and deterministic writing and deleting process. Using micromagnetic simulations, we also reveal the microscopic mechanism of the topological fluctuations that occur during this process.**

Nanoscale memory devices that employ magnetic materials and offer non-volatility and efficient electrical controllability have the potential to outperform and eventually replace conventional silicon-based memory technologies such as DRAM or NAND-Flash[1–3]. Among the many different types of spintronic memory devices available, those based on magnetic skyrmions – topologically non-trivial spin nanostructures – are of particular interest.[4–10] Key advantages of skyrmion-based memory device include their stability[9], high-density arrangement[4,5] and low-power operation using electrical excitations[6–8].

Recently the room-temperature stabilization of magnetic skyrmions and their current pulse-induced displacement on nanotracks has been reported in magnetic heterostructures such as Ta/CoFeB/MgO[11,12], Pt/Co/Ta[13], Pt/CoFeB/MgO[13–15], Pt/Co/Ir[16] and Pt/GdFeCo/MgO[17] where the strong interfacial Dzyaloshinskii-Moriya interaction (DMI) leads to the stabilization of the skyrmion structures[18,19]. While such current-controlled skyrmion motion is applicable to an actual device scheme, the deterministic writing and deleting of a single isolated magnetic skyrmion at room temperature is required for fully functional skyrmionic devices[10,20].

Theoretical studies have suggested that skyrmions can be created by applying localized vertical spin-polarized current to a confined geometry[21] or well-defined notches[22]. Moreover, there are suggestions that a single skyrmion can be formed by the conversion of a pair of domain walls[23,24]. However, the controlled creation of a single skyrmion, and the subsequent annihilation of the same skyrmion, has only been experimentally achieved at low temperatures with spin-polarized scanning tunneling microscopy (SP-STM)[8,25]. This approach requires the in situ deposition of material at cryogenic conditions, which may be hard to incorporate into actual commercial devices. Recently, several studies have experimentally reported the electrical creation of single or multiple magnetic skyrmions at room temperature in metallic ferromagnetic heterostructures[16,26–28], which could be applicable to a skyrmionic device scheme. However, the electrically controlled room-temperature annihilation of a single magnetic skyrmion has so far only been accessible with micromagnetics[29–31]. Moreover, it will be essential to find a scheme and material system where both topological transitions – creation and annihilation – are readily available.

In this Article, we demonstrate the electrical creation and annihilation of a single isolated magnetic skyrmion at room temperature in a practical device scheme. To reveal the deterministic and reproducible nature of this process, we use stroboscopic pump-probe X-ray measurements, which are complemented by micromagnetic simulations. Our work provides an electrical manipulation method that should be directly applicable to any skyrmion-based device application.

**Sample structure and static X-ray microscopy measurement**

[Pt (3 nm)/Gd$_{25}$Fe$_{65.6}$Co$_{9.4}$ (7 nm)/MgO (1 nm)]$_{20}$ multilayer stacks (hereafter Pt/GdFeCo/MgO) based on ferrimagnetic Gd$_{25}$Fe$_{65.6}$Co$_{9.4}$ with perpendicular magnetic anisotropy (PMA) were studied (see Methods for the details). The ferrimagnetic multilayer structure was chosen due to the efficient current pulse-driven skyrmion motion with significantly reduced skyrmion Hall angle achieved in the system[17]. Figure 1 shows the magnetic property of our multilayer structure as well as the magnetic field- and electric current-driven evolution of magnetic configurations in the material observed by static X-ray measurements. Figure 1a first shows that the out-of-plane hysteresis loop of the Pt/GdFeCo/MgO multilayer has no remnant magnetization due to the strong demagnetization field, resulting in labyrinth multi-domain configuration at a zero magnetic field (see the first image of Fig. 1b). Figure 1b presents a series of scanning transmission X-ray microscopy (STXM) images of the Pt/GdFeCo/MgO multilayer patterned into magnetic tracks acquired at different magnetic fields. Note that every STXM image shown in this article was taken at the absorption edge of Fe atom ($L_3$-edge: 708 eV). The bright and dark contrasts in STXM images correspond to downward (-$M_z$) and upward (+$M_z$) out-of-plane magnetization directions of Fe atoms, respectively. Due to the antiferromagnetic spin ordering between Fe and Gd within the GdFeCo alloy and the Gd magnetic moment being larger than the Fe magnetic moment at $T_{RT} < T_{M-compensation}$, the magnetic moment of Gd atoms aligns parallel with the external magnetic field while the magnetic moment of Fe atoms aligns anti-parallel to the Gd magnetic moments[17,32].

Upon the magnetic field application in the +$z$ direction, the initial labyrinth domain state (at $B_z$ = 0 mT) transforms into a multi-bubble domain state (at $B_z$ = +110 mT), and eventually, a uniform magnetization state is achieved at a larger field (at $B_z$ = +130 mT). To determine the chiral nature of magnetic bubbles observed at $B_z$ = +110 mT in Fig. 1b, we quantified the electric current pulse-driven behaviors of the bubble domains as shown in Fig. 1c. Each STXM image in Fig. 1c was acquired after injecting current pulses. Between two adjacent STXM images, 20 repetition of the current pulses, with a current density of $|j_a|$ = 4.23×10$^{11}$ A m$^{-2}$ and pulse duration of 5 ns, were applied. The application of current pulses into the Pt/GdFeCo/MgO multilayer structure is expected to excite spin orbit torques (SOTs)[33,34] originating at the interface between Pt and GdFeCo. From the series of acquired STXM images, we find that two of the nine initial bubbles move along the direction of current flow. Note that, upon the application of SOTs, only chiral skyrmions can be displaced while non-chiral bubbles can only be either expanded or annihilated[11]. Moreover, depending on the type of chirality, chiral skyrmions move along either electron-flow (for right-handedness) or current-flow (for left-handedness) directions. Therefore, observations shown in Fig. 1c confirm that our bubbles are left-handed Néel-type chiral skyrmions. We also observe the presence of relatively strong pinning sites in this material, which pinned seven out of the nine skyrmions during current pulse injections. These observations agree with our previous observation in the same multilayer system[17]. Figure 1d shows the schematic drawing of a ferrimagnetic skyrmion observed in Fig. 1c. With the above basic information on our material system and observed skyrmions, we now demonstrate single skyrmion writing-deleting using a pump-probe technique applied to the STXM measurements.

**Time-resolved measurements on skyrmion writing and deleting**

Figure 2 summarizes the results of our pump-probe experiments. Figure 2a,b first show the schematic drawing of our pump-probe experiment and the pulse profiles used for the electrical excitation, respectively. A representative STXM image and the actual sample image acquired using scanning electron microscopy (SEM) are also enclosed in Fig. 2a. From the SEM image, we find that there exists material residue near the wire edges due to the incomplete lithography, and this resulted in the lateral contrast gradient in Fig. 2e,f. Figure 2b shows the pulse profile used for the pump-probe experiments. Two 10 ns current pulses of opposite polarity were applied at 1000ns intervals. The relatively strong reflected pulses due to the presence of sizable impedance mismatch ($R_{sample} \approx 300$ Ohms), make the two pulses of opposite polarity in effect two bipolar pulses of $|j_2|$ (amplitude of the reflected pulse) $\approx 0.2$ $|j_1|$ (amplitude of the initial pulse). We applied a constant field, $B_z = +130$ mT, to drive the magnetization into uniformly magnetized state, and this state serves as the initial magnetic configuration of the following current pulse driven dynamics measurements.

The time-resolved X-ray microscopy utilized the synchronized current pulses injections (pump) and the X-ray flashes of ~50 ps temporal width at a 2 ns repetition rate (probe). Due to the relatively low intensity for a single X-ray flash, the averaged image of roughly $10^{10}$ flashes was used for dynamic sequence. This important characteristic of the pump-probe technique implies that only fully repeatable and deterministic events can be imaged. Figure 2c,d first show the profile of bipolar pulse, labelled as type-I through this article, (writing pulse, Fig. 2c) and the reversed type-I pulse (deleting pulse, Fig. 2d) injected into the GdFeCo magnetic track. The time-resolved STXM images recorded during these pulse applications are shown in Figs. 2e,f. The coloured circles in Fig. 2e,f indicate the time delay corresponding to those of the same colour in Figs. 2c,d.

Figure 2e shows the time-dependent evolution of magnetic textures during the writing pulse. The initial state (image #1) shows no visible magnetic texture in the track. However, as soon as the pulse turns on, up to seven 100~200 nm-diameter circular reverse domains are nucleated and remain stable (images #2-#4), which are acknowledged as skyrmions based on our prior current driven translation results in Fig. 1c. Note that the conventional SOT-induced magnetization reversal requires a large in-plane magnetic field, typically $B_z \geq +100$ mT, while the effective in-plane magnetic field applied to our sample was measured to be only $|B_{x\text{-exp.}}| \approx +5$ mT, which may not be sufficient for an effective magnetization canting. However, earlier studies show that the observed magnetization reversal is possible due to the strong DMI within a magnetic film of non-uniform PMA, $K_u$[28]. Therefore, the areas, where reversed domains are nucleated, may be the regions of locally reduced anisotropy. Although regions of reduced anisotropy were not intentionally fabricated, it is plausible that there could be spatial variation of the magnetic anisotropy, e.g. pinning sites with reduced anisotropy, within the sputter deposited multilayers. To the state shown in image #4, we then switched the pulse direction (images #5-#8), and surprisingly, we observe only two skyrmions are left and the other skyrmions have disappeared. Eventually, as the pulse turns off, only one ~150 nm-sized skyrmion remains present. The observed time-resolved dynamics demonstrates the detailed evolution of magnetic configuration during a current pulse induced single skyrmion writing process. It is noteworthy that the written skyrmion may still be located near a pinning site that served as a skyrmion generation source; therefore, it will be required to displace the written skyrmion using electrical pulses for practical device applications. As presented in Supplementary Fig. 1 and Supplementary Note 1, we have demonstrated that the electrically written skyrmion can be freed from the pinning site and displaced along a magnetic nanotrack by simply modulating the external magnetic field and electric pulses.

To the single skyrmion state (shown in image #8), we reverse the pulse polarity and apply the deleting pulse along the stripe. Figure 2f shows the time-dependent evolution of magnetic textures during the deleting pulse. While the single skyrmion state shows no observable changes during the rising time, as soon as the pulse reaches its maximum and flattens (image #13), the skyrmion becomes annihilated and the uniform magnetization state is achieved (images #13-#16). The full movie of the single skyrmion (labelled as $Sk_2$ in Fig. 2e) writing/deleting can be found in Supplementary Video 1. It is noteworthy that, in Supplementary Video 1, other skyrmions except $Sk_2$ are hard to be noticed because they only exist during 5 frames (corresponding to 10 ns as presented in Fig. 2e) out of 1000 total video frames (corresponding to 2000 ns). Therefore, for the improved visibility on the skyrmion writing and deleting, we provide the writing and deleting parts as separate videos: Supplementary Video 2 for skyrmion writing and Supplementary Video 3 for skyrmion deleting. The significance of the observation is that the observed creation and annihilation of a single skyrmion are deterministic and reproducible due to the nature of measurement scheme: pump-probe imaging records averaged images of roughly $10^{10}$ measurements, thus, all magnetic structures at each time delay should be seen reproducibly in the vast majority of integration cycles, otherwise, significant loss in magnetic contrast will make them difficult to be observed. In Supplementary Figs. 2-3 and Supplementary Note 2, we demonstrate that writing and deleting of multiple skyrmions are also achievable by simply changing the pulse amplitude, which alters the magnitude of SOTs.

**Micromagnetic simulation on skyrmion writing**

Having experimentally established that skyrmions can be written and deleted electrically in this material, we study the microscopic origin of the observed phenomena using spin dynamics simulations. Figure 3a shows a $400 \times 400$ nm$^2$ effective ferrimagnetic layer used for simulations (see Methods), and the example of defect-like site is realized as the region of reduced PMA (Fig. 3b). This type of pinning potential landscape is easily expected for sputter-grown non-uniform interface structures and has been extensively studied in previous reports[12,13,35]. To the initial state, we apply the current pulses with modified profiles as shown in Fig. 3c (type-I pulse) and 3d (type-II pulse). Type-I pulse imitates the pulse shape acquired in our experiments and uses the same label while the type-II pulse is constructed based on the recent report by Büttner et al[28]. Figure 3c,d plot the variation of topological charge $Q$ (see Methods) in each sub-lattice during the pulse applications. Note the sub-lattice topological charge $Q$ is defined and calculated separately on each sub-lattice. During the application of type-I pulse, mere topology fluctuation during the strong downward pulse becomes significantly amplified perturbation during the following upward pulse, eventually writing a single magnetic skyrmion with $|Q| = 1$ (see Supplementary Video 4 and Supplementary Video 5), and this observation is in agreement with experimental observation shown in Fig. 2e. However, for the application of type-II pulse, the topological charge of each sub-lattice only fluctuates without noticeable changes. While both pulse types create a bubble with $Q = 0$ by the microscopic process of SOT-induced magnetization switching, for the case of type-II pulse, the created bubble structure shrinks and becomes annihilated as soon as the pulse turns off due to the dipolar field effect and its topologically trivial nanoscale structure (see Supplementary Video 6). Figure 3g illustrates the time evolution of the spin configuration during the skyrmion writing achieved in Fig. 3c, where the same process is schematically drawn in Fig. 3h for better comprehension over the process.

It can be seen that when the strong downward pulse is applied, the magnetic moments in the reduced anisotropy region reverse at first (at $t = 120$ ps), where a bubble with $Q = 0$ is generated and the bubble contains a localized vertical Bloch line (VBL). For the remaining pulse (at $t = 200$ ps), due to the driving force provided by the SOTs, the two parts of the reverse domain carrying opposite topological charges will move in an opposite direction, leading to the expansion of the reverse domain. However, following small pulse with opposite polarity reverses the motion directions of both the Gd and FeCo domains carrying opposite topological charges, which forces the bubble to shrink. This process will result in the destruction of the defect-like point (VBL) and the formation of a topologically non-trivial structure, i.e. a skyrmion with $|Q| = 1$. Thus, the application of the weak upward pulse, which eventually forces to annihilate the VBL and turns the trivial common bubble into a non-trivial chiral skyrmion, plays a crucial role for the writing process.

On the other hand, for the type-II unipolar pulse, the first large pulse creates a bubble with sublattice topological charge of zero, $Q = 0$, as the bubble contains a defect-like point, VBL. Under the driving force provided by the SOT, domain walls carrying opposite topological charges move in opposite directions. Therefore, the size of the reverse domain increases under the larger pulse. The following small pulse with the same polarity continues to increase the reverse domain size and drive it into motion. When the pulse is turned off, the reverse domain shrinks, which can result in the destruction of the VBL and the formation of a skyrmion. This is the creation process of a skyrmion described in Ref. [28]. From above discussions, a conclusion can be drawn that the shrinking of the reverse domain leads to the destruction of the VBL and thus generates a skyrmion. However, spontaneous shrinking may not always destroy the VBL.

Therefore, we then compare the efficiency of two pulse types as a function of the amplitude and width of $j_2$ (secondary weak pulse), and the resulting phase diagrams are shown in Figs. 3e,f. It is clear that the type-I pulse provides very wide available area in the phase diagram (Fig. 3e) while the type-II pulse offers a very narrow range of pulse width and amplitude (Fig. 3f) that are able to create the skyrmion structure. Moreover, the ultrafast creation of a single skyrmion in the time scale of ~50 ps is achievable with the type-I pulse. Therefore, we find that the bipolar pulse with unbalanced amplitude, as used in experiments and labeled as type-I pulse in this study, offers an efficient and practical method for the skyrmion writing. We also numerically compared the efficiency of two pulses as a function of the amplitude and width of $j_1$ (primary strong pulse) in Supplementary Fig. 4, and confirmed far enhanced performance of type-I pulse for skyrmion generation. Indeed, as studied in Refs. [21,22,28], the creation of the skyrmion also depends on the condition of the defect-like site. We show in Supplementary Fig. 5, Supplementary Note 3, and Supplementary Video 7 that bubbles can be simultaneously created from several different defect-like sites during the application of the type-I pulse, however, not all bubbles can be transformed to skyrmions, which agrees with our experimental observations. Moreover, we reveal in Supplementary Fig. 6 that, either driven by the type-I or type-II pulse, a defect-like site with larger diameter or smaller PMA serves as a more efficient source for the creation of skyrmion.

**Micromagnetic simulation on skyrmion deleting**

Skyrmion deletion upon pulse application is also investigated by simulations. During the skyrmion deletion simulations, a finite in-plane field of $|B_{x\text{-simul.}}| = +70$ mT was applied, considering the experimentally measured small in-plane field, $|B_{x\text{-exp.}}| = +5$ mT. The temperature effect is ignored in our zero-temperature simulations. We confirm that the application of this in-plane field breaks the symmetry and results in the efficient skyrmion deletion via pulse injections. From the experimental results shown in Figs. 2d,f, we have observed that skyrmion deletion occurs as soon as the pulse turns on, therefore, we simulated the annihilation process of a single skyrmion using a simple unipolar pulse with varying pulse lengths and amplitudes. The model is a 400 × 400 nm$^2$ square sample with ferrimagnetic ordering at a constant background field of $B_z = +130$ mT, and other parameters are given in the Methods section. At the beginning of the simulation, a relaxed ferrimagnetic skyrmion is placed at the centre of the sample, and we apply a unipolar pulse to delete the skyrmion. The current density of the pulse is in the range of $j = 1.5 \times 10^{12} \sim 3.0 \times 10^{12}$ A m$^{-2}$, and the pulse length is in the range of $\tau = 0 \sim 200$ ps. After the pulse injection, the sample is relaxed to the stable/metastable state.

As summarized in Fig. 4a, we have identified that five different topological fluctuation regimes exist, denoted as Cases I-V in Fig. 4a, depending on the external pulse conditions. Time-dependent snapshots of magnetic configurations are also shown in Fig. 4b (Case II and Case V, where skyrmion deletion is observed) and Supplementary Fig. 7 (Case I, Case III and Case IV, where skyrmion deletion is not observed). In Case I, the pulse has of weak-amplitude and short-duration, leading to the negligible topological changes and the skyrmion remains stable. In Case II, where a small-amplitude pulse is applied for a long time, the skyrmion is pushed to the edge and destroyed as shown in Fig. 4b. Although the pulse amplitude was relatively small to cause the internal skyrmion collapse solely driven by the pulse, the long pulse duration could drive the skyrmion to near the boundary, leading to the strong interaction between the skyrmion and the boundary that eventually resulted in the skyrmion collapse. However, it should be noted that this mechanism only occurs within a relatively narrow range of pulse duration and amplitude, as shown in Fig. 4a. The full process of skyrmion deletion in Case II can be found in Supplementary Video 8.

In Case III-V, we observed strong internal domain fluctuations as shown in Fig. 4 and Supplementary Fig. 7. However, the transformation develops into a typical bubble with a topological defect, VBL, only in Case V (see Fig. 4b), where a large-amplitude and long-duration pulse was applied. Then, the topologically unprotected bubble eventually shrinks and collapses due to a dipolar field. Figure 4c presents the evolution of skyrmion number $Q$ of each sub-lattice of ferrimagnetic structure and current pulses as a function of simulation time for Case V. It presents that, as soon as pulse turns on, internal domain fluctuation initiates within ~ 50 ps, which eventually generates a typical bubble with $Q = 0$ as shown in Fig. 4b and Supplementary Video 9. Unlike the Case II, this skyrmion deletion occurs internally without interacting with the boundary, as was also observed using micromagnetic simulations in Refs. [29–31]. This mechanism offers a very wide range of pulse width and amplitude (Fig. 4a) that are able to delete the skyrmion structure. From the presented deletion process of Case V, we again confirm that the appearance of the topological defect, which can be excited by electrical pulse, plays a crucial role for the deleting process. Although we could not experimentally confirm the strong internal fluctuation and the injection of VBL as observed in simulations, mainly due to the limited temporal and spatial resolution, we expect that the experimentally observed skyrmion deletion initiated by 10 ns-long pulse may have gone through this type of internal annihilation process. We would like to note that skyrmion deletion always occurs via Case V as long as the strong pulse is applied longer than roughly ~80 ps, offering a very wide range in the skyrmion deletion phase diagram (Fig. 4a).

Using simulations, we further reveal that the detailed internal domain fluctuation could vary depending on the local variation of other parameters, such as DMI, as shown in Supplementary Fig. 8, or the mesh size, as shown in Supplementary Fig. 9. The simulation results point that the internal skyrmion deletion process should mediate the pulse-induced generation of topological defect (VBL), which may have occurred in experiments. Overall, our simulations reproduce the experimental observations presented in Fig. 2 with great qualitative agreement, and further reveal the current pulse-dependent skyrmion deleting mechanisms that are readily accessible by adjusting electrical pulse profiles.

**Conclusion**

Using time-resolved X-ray pump-probe measurements, we have demonstrated the creation and deletion of a single isolated magnetic skyrmion in ferrimagnetic GdFeCo films. The microscopic origin behind the observed time-dependent magnetic texture evolution was also elucidated using spin dynamics simulations. We have shown that regions with locally reduced anisotropy can serve as local skyrmion sources and provided a current pulse profile that offers a technological route toward creating a skyrmion, which can further be displaced along a magnetic track by additional current pulse injections. We also showed that by reversing the polarity of the writing-pulse, we can efficiently delete the generated skyrmion, and using micromagnetic simulations, reveal the distinct electric current pulse induced skyrmion deleting mechanisms.

Our approach shows that the creation and subsequent annihilation of the same single skyrmion can be easily achieved using electrical methods on a single magnetic device with a technologically-relevant thin-film racetrack geometry; a geometry in which the current-driven displacement of a train of individual skyrmions at the speed approaching 50 m s$^{-1}$ has already been demonstrated.[17] Several techniques on the electrical manipulation of magnetic skyrmions have very recently been reported[16,27,28,36]. We believe that, together with these techniques, our findings could help deliver fully functional skyrmion-based device applications at room temperature.

**Methods**

**Experimental details.** The [Pt(3 nm)/GdFeCo(7 nm)/MgO(1 nm)]$_{20}$ films were grown by DC magnetron sputtering at room temperature under 1 mTorr Ar for Pt and GdFeCo and 4 mTorr Ar for MgO at a base pressure of roughly ~3×10$^{-8}$ Torr. As illustrated in Fig. 2a, a 500 nm-wide and 5 μm-long magnetic stripe, consists of Pt/GdFeCo/MgO multilayer, was patterned on a 100 nm-thick Si$_3$N$_4$ membrane using electron beam lithography and lift-off technique. The two side edges of the stripe were connected with Ti(5 nm)/Au(100 nm) contacts for current applications. Nominally same films were grown on thermally oxidized Si substrate for vibrating-sample magnetometry (VSM) measurement, and the measurement yielded material constants of $\mu_0 H_k$ = 0.15 T, and a net saturation magnetization $M_S$ = 2×10$^5$ A m$^{-1}$. All microscopy images were acquired using the STXM installed at the MAXYMUS endstation of the BESSY II, HZB (Berlin, Germany). The devices used for the experiments were 5-μm-wide and 7-μm-long (for static measurements shown in Fig. 1) and 0.5-μm-wide and 7-μm-long (for dynamics measurements shown in Fig. 2 and Supplementary Videos 1-3). Note that the wire structure used for pump-probe experiments yielded an electrical resistance of $R_{sample}$ ~ 300 Ohms measured in 2-point. The excitation pulses were simultaneously verified before and after the sample through two -20 dB pick-off tees that show pick-off signals to an oscilloscope. To minimize the effect of Joule heating, we maintained low duty-cycle, ~1%, during dynamics measurement and supplied the He flow of 10 mTorr for active cooling. To improve the visibility of the raw images acquired from X-ray measurements, we performed contrast/brightness adjustments and enhanced sharpness of the magnetic objects to better demonstrate detailed evolution of magnetic skyrmions. We would also like to note that the adjustment has not changed important physical characteristics of observed skyrmions, e.g. skyrmion diameter.

**Simulation details.** The spin dynamics simulation is performed by using the 1.2a5 release of the Object Oriented MicroMagnetic Framework (OOMMF)[37]. The model is treated as a checkerboard-like two-sub-lattice spin system based on the G-type antiferromagnetic structure with simple square lattices[38], where the two sub-lattices are coupled in a ferrimagnetic manner with a net spontaneous magnetization, while each sub-lattice is ferromagnetically ordered. The spin Hamiltonian $\mathcal{H}$ is given by the expression

$$\mathcal{H} = -J_{ij} \sum_{<i,j>} \mathbf{S}_i \cdot \mathbf{S}_j + D \sum_{<i,j>} (\mathbf{u}_{ij} \times \hat{z}) \cdot (\mathbf{S}_i \times \mathbf{S}_j) - K \sum_i (S_i^z)^2 - \sum_i \mu_i \mathbf{S}_i \cdot \mathbf{H} + H_{DDI}, \quad (1)$$

where $\mathbf{S}_i$ represents the local spin vector reduced as $\mathbf{S}_i = \mathbf{M}_i/M_S^i$ at the site $i$, and $\mathbf{S}_j$ represents the local spin vector reduced as $\mathbf{S}_j = \mathbf{M}_j/M_S^j$ at the site $j$. $\mathbf{M}_i$ and $\mathbf{M}_j$ are the magnetization at the site $i$ and $j$, respectively. $M_S^i$ denotes the saturation magnetization of the sub-lattice $i$, while the saturation magnetization of sub-lattice $j$ is defined as $M_S^j = nM_S^i$ with the compensation ratio $n$. $<i,j>$ runs over all the nearest-neighbor sites in the two-sub-lattice spin system. $J_{ij}$ is the exchange coupling energy constant between the two spin vector $\mathbf{S}_i$ and $\mathbf{S}_j$, which has a negative value ($J_{ij} < 0$) representing the antiferromagnetic spin ordering of the two sublattices. $D$ is the interface-induced DMI constant, $\mathbf{u}_{ij}$ is the unit vector between spins $\mathbf{S}_i$ and $\mathbf{S}_j$, and $\hat{z}$ is the interface normal, oriented from the heavy-metal layer to the ferrimagnetic layer. $K$ is the PMA constant, $\mathbf{H}$ is the applied magnetic field, $\mu_i$ is the magnetic moment of the site $i$, and $H_{\text{DDI}}$ stands for the dipolar interactions.

The time-dependent spin dynamics is controlled by the Landau-Lifshitz-Gilbert (LLG) equation augmented with the damping-like spin Hall torque provided by electrons that flow through the Pt layer, which is expressed as

$$\frac{d\mathbf{S}_i}{dt} = -\gamma_0 \mathbf{S}_i \times \mathbf{H}_{\text{eff}} + \alpha\left(\mathbf{S}_i \times \frac{d\mathbf{S}_i}{dt}\right) + \tau[\mathbf{S}_i \times (\hat{p} \times \mathbf{S}_i)], \qquad (2)$$

where $\mathbf{H}_{\text{eff}} = -(1/\mu_0 M_S^i) \cdot (\delta\mathcal{H}/\delta\mathbf{S}_i)$ is the effective field on a lattice site, $\gamma_0$ is the Gilbert gyromagnetic ratio, and $\alpha$ is the phenomenological damping coefficient. Note we assume $\gamma_0$ and $\alpha$ are identical for the two sub-lattices. The coefficient for the spin Hall torque is given as $\tau = (\gamma_0 \hbar j \theta_{\text{SH}})/(2\mu_0 e M_S^i b)$, where $j$ is the applied charge current density, $\theta_{\text{SH}}$ is the spin Hall angle, and $b$ is the thickness of the ferrimagnetic layer. $\hat{p} = \mathbf{j} \times \hat{\mathbf{z}}$ denotes the spin polarization direction.

For the simulation on the multilayer structure, we employed an effective medium approach[13], which improves the computational speed by converting the multilayer into a two-dimensional effective model with reduced parameters. In the effective medium model, the thickness of one ferrimagnetic layer is $t_m$ = 7 nm, the thickness of one repetition is $t_r$ = 11 nm, the number of repetitions is $n_{\text{rep}}$ = 20, and the lateral cell size is $2 \times 2$ nm$^2$. The open boundary condition is applied in the simulation. The intrinsic magnetic parameters used in the simulation are measured from our experimental samples as well as adopted from Ref. [17]: the Gilbert damping coefficient $\alpha$ = 0.5, the inter-sub-lattice exchange stiffness $A$ = -15 pJ m$^{-1}$, the spin Hall angle $\theta_{\text{SH}}$ = 0.1, the DMI constant $D$ = -1.5 mJ m$^{-2}$, the PMA constant $K_u$ = 0.04 MJ m$^{-3}$, and the net saturation magnetization $M_S = M_S^{\text{Gd}} - M_S^{\text{Fe}}$ = 200 kA m$^{-1}$. The compensation ratio is defined as $n = M_S^{\text{Fe}}/M_S^{\text{Gd}} = 0.7$. The effective spin Hall angle and DMI constant are reasonably acquired from Pt-based ferromagnetic heterostructures in Refs. [13,15,39]. For the simulation of skyrmion writing phase diagram, the default pinning site profile is given in Fig. 3b, where the diameter equals 150 nm and the minimal anisotropy at the pining centre equals $0.1K_u$.

Besides, both the bubble and skyrmion are characterized by the topological charge on either sub-lattice, which is given as

$$Q_i = -\frac{1}{4\pi} \int \mathbf{S}_i \cdot \left(\frac{\partial \mathbf{S}_i}{\partial x} \times \frac{\partial \mathbf{S}_i}{\partial y}\right) dxdy. \tag{3}$$

The topological charge $Q$ is also referred to as the skyrmion number. A topologically trivial ferrimagnetic bubble has $Q = 0$ on both sub-lattices, while a topologically non-trivial ferrimagnetic skyrmion has $Q = +1$ on sub-lattice $i$ and $Q = -1$ on sub-lattice $j$, or vice versa.

**Data Availability Statement**

The data that support the plots within this paper and other findings of this study are available from the corresponding author upon reasonable request.

**Acknowledgements**

This work was primarily supported by Samsung Research Funding Center of Samsung Electronics under Project Number SRFC-MA1602-01. Most experiments were performed at the MAXYMUS endstation at BESSY2, HZB (Berlin, Germany). The authors acknowledge HZB for the allocation of beamtime. S.W., M.-C.P., K.-Y.L., J.W.C., B.-C.M., H.C.K. and J.C. acknowledges the support from KIST Institutional Program. K.M.S acknowledges the support from the Sookmyung Women's University BK21 Plus Scholarship. X.Z. was supported by JSPS RONPAKU (Dissertation Ph.D.) Program. Y.Z. acknowledges the support by the President's Fund of CUHKSZ, the National Natural Science Foundation of China (Grant No. 11574137), and Shenzhen Fundamental Research Fund (Grant Nos. JCYJ20160331164412545 and JCYJ20170410171958839). M.E. acknowledges the support by the Grants-in-Aid for Scientific Research from JSPS KAKENHI (Grant Nos. JP17K05490, 25400317 and 15H05854), and also the support by CREST, JST (Grant No. JPMJCR16F1). J.W.C. acknowledges the travel fund supported by the National Research Foundation of Korea (NRF) funded by the MSIP (2016K1A3A7A09005418). S.W. and B.-C.M. acknowledge the support from the National Research Council of Science & Technology (NST) grant (No. CAP-16-01-KIST) by the Korea government (MSIP).

**Author Contributions**

S.W. designed and initiated the study. K.M.S. performed film growth, optimization and device fabrication on $Si_3N_4$ membranes. S.W., K.M.S. and J.W.C performed X-ray experiments with supports from M.W. using STXM at BESSY2 in Berlin, Germany. X.Z., Y.Z. and X.L. performed the numerical simulations. M.E. carried out the theoretical analysis. During the revision of this article, S.W., K.M.S., S.F. and J.R. performed X-ray imaging experiments as summarized in Supplementary Fig. 1 using STXM at Swiss Light Source in Villigen, Switzerland. M.-C.P. and K.-Y.L. provided technical supports. S.W. and X.Z. drafted the manuscript and revised it with assistance from M.-C.P., K.-Y.L., J.W.C., B.-C.M., H.C.K. and J.C.. All authors commented on the manuscript.

**Competing Interests**

The authors declare no competing interests.

**Author Information**

**Figure Legends**

**Figure 1. Hysteresis behavior of ferrimagnetic multilayer structure, and field- and current-driven domain evolution within the structure. a,** Out-of-plane hysteresis loop for [Pt/GdFeCo/MgO]$_{20}$ multilayer structure measured by vibrating sample magnetometry (VSM). Inset shows out-of-plane hysteresis loop for [Pt/GdFeCo/MgO]$_1$ unit structure, showing square-shaped easy-axis loop. A series of scanning transmission X-ray microscopy (STXM) images acquired from ferrimagnetic stripe structure in the presence of **b,** magnetic field, from $B_z = 0$ mT to $B_z = +130$ mT, and **c,** electric current pulses, where $|j_a| = 4.23 \times 10^{11}$ A m$^{-2}$. Note that the dark and bright contrasts in STXM images correspond to upward ($+M_z$) and downward ($-M_z$) out-of-plane magnetization directions of Fe atoms, respectively. During these static measurements, each image was obtained after magnetic field- or current pulse-application. For current pulse-driven skyrmion behavior measurement, mobile skyrmions are highlighted and tracked as indicated with green and red colors in **c**. **d,** Schematic drawing of a ferrimagnetic skyrmion observed in **c**. The scale bar is 2 μm.

**Figure 2. Time-resolved pump-probe X-ray microscopy measurement. a,** Schematic drawing of a device used for time-resolved scanning transmission X-ray microscopy (STXM) measurements. An exemplary STXM image and a scanning electron microscope (SEM) image of the actual sample are enclosed. During pump-probe measurements, each X-ray photon flash, which has roughly 50 ps of effective width, is injected at the frequency of $f_{\text{X-ray}} = 499.65$ MHz. The scale bar is 1 μm. **b,** The pulse cycle used for the pump-probe dynamics measurement. Two 10 ns-long pulses were injected with 1 μs interval, and as discussed in the main text, two pulses resemble two bipolar pulses of $|j_2|$ (amplitude of reflected pulse) ≈ 0.2 $|j_1|$ (amplitude of the initial pulse) due to a large sample resistance. The pulse polarity represents the direction of current flow $j_a$. For the positive polarity, electric current flows from the Au electrode on the left to that on the right. Detailed (reversed) type-I pulse profiles used for **c,** skyrmion writing and **d,** skyrmion deleting dynamics measurements. The colored circles in this plot indicate the time delay, which are also shown in following STXM images. **e, f,** Magnetic skyrmion configuration at different time delays at the voltage amplitudes of $V_a = 3$ V, corresponding to the current density of $j_a = 2.5 \times 10^{10}$ A m$^{-2}$. Constant out-of-plane field of $B_z = +130$ mT was applied during the electrical excitation.

Time numbers, #1-#16, are included for convenience, and noticeable magnetic skyrmions (regardless of their non-chiral nature) are individually labeled. Scale bar is 500 nm.

**Figure 3**. **Micromagnetic simulation of the single skyrmion writing. a,** The initial ferrimagnetic ordering of a 400×400 nm$^2$ square sample at $B_z$ = +130 mT used for simulations. Central circular region indicates the area with reduced out-of-plane magnetic anisotropy. Scale bar is 100 nm. **b,** The spatial profile of the out-of-plane magnetic anisotropy within the circular region in **a**. **c, d,** Skyrmion number ($Q$) of each sub-lattice of ferrimagnetic structure and applied current pulses as a function of simulation time. The ground state topological charge is not exactly zero ($Q \neq 0$) due to the canted spin orientation near mesh boundaries. Sub-lattice A and B correspond to Gd and Fe, respectively, and type-I and type-II pulses are shown in **c** and **d**, respectively. Type-I consists of a 200 ps-long strong pulse ($j_1$ = -1.5×10$^{12}$ A m$^{-2}$) followed by another 200 ps-long weak pulse ($j_2$ = -0.6$j_1$) of the opposite polarity while the weak pulse of the same polarity ($j_2$ = 0.6$j_1$) is used for type-II. Note that the type-I pulse imitates the detailed pulse profile used for experiments as shown in Fig. 2. The current density is larger than the experimental value because, in real experiments, the temperature effect may result in smaller critical current density for effective topological fluctuation. **e, f,** Skyrmion writing phase diagram in which the skyrmion number $Q$ after current pulse injection and relaxation was determined for **e**, type-I and **f**, type-II, respectively. Phase diagrams are drawn as functions of the amplitude and pulse width $\tau$ of $j_2$ at a fixed 100 ps-long $j_1$ ($j_1$ = -1.7×10$^{12}$ A m$^{-2}$). The model for the phase diagram simulation is optimized to an 800×400 nm$^2$ rectangular sample. **g,** Time-dependent evolution of skyrmion during the application of type-I pulse. Vertical Bloch lines (VBLs) are indicated with a yellow circle. Scale bar is 100 nm. The full movie of this skyrmion writing process can be found in Supplementary Video 4 and Supplementary Video 5. **h,** Schematics of the time-dependent process of a single skyrmion generation.

**Figure 4**. **Micromagnetic Simulation of the single skyrmion deletion. a,** Phase diagram showing different regimes of the skyrmion deletion depending on the amplitude ($j$) and length ($\tau$) of the current pulse. Background in-plane magnetic field of $B_x$ = -70 mT was applied. Case I: the skyrmion is not destroyed by the pulse and remains stable after the pulse injection. Case II: the skyrmion is pushed to the edge and destroyed driven by the pulse. Case III: the skyrmion is pushed to the edge and destroyed driven by the pulse; meanwhile, the pulse

creates a new skyrmion. Case IV: the skyrmion is destroyed and transformed to domain walls driven by the pulse in the interior of the sample, and a new skyrmion is created from the domain wall after the pulse injection. Case V: the skyrmion is destroyed and transformed to domain walls driven by the pulse in the interior of the sample, and the sample is relaxed to the ferrimagnetic state after the pulse injection. **b,** Time-dependent snapshots of magnetic configurations showing the skyrmion deletion process corresponding to different cases given in **a**. Scale bar is 100 nm. The parameters for each case are: Case II, $j = 1.6\times10^{12}$ A m$^{-2}$, $\tau = 185$ ps; Case V, $j = 2.2\times10^{12}$ A m$^{-2}$, $\tau = 170$ ps. As described in the main text, the skyrmion in Case II is deleted at the sample edge while the skyrmion in Case V is deleted at the interior of the sample, which shows good agreement with experimental observation. Vertical Bloch lines (VBLs) are indicated with a yellow circle. Full movie of each deletion process can be found in Supplementary Video 8 (Case II) and Supplementary Video 9 (Case V). **c,** Skyrmion number ($Q$) of each sub-lattice of ferrimagnetic structure and applied current pulses as a function of simulation time for Case V, which is the interior skyrmion deletion.

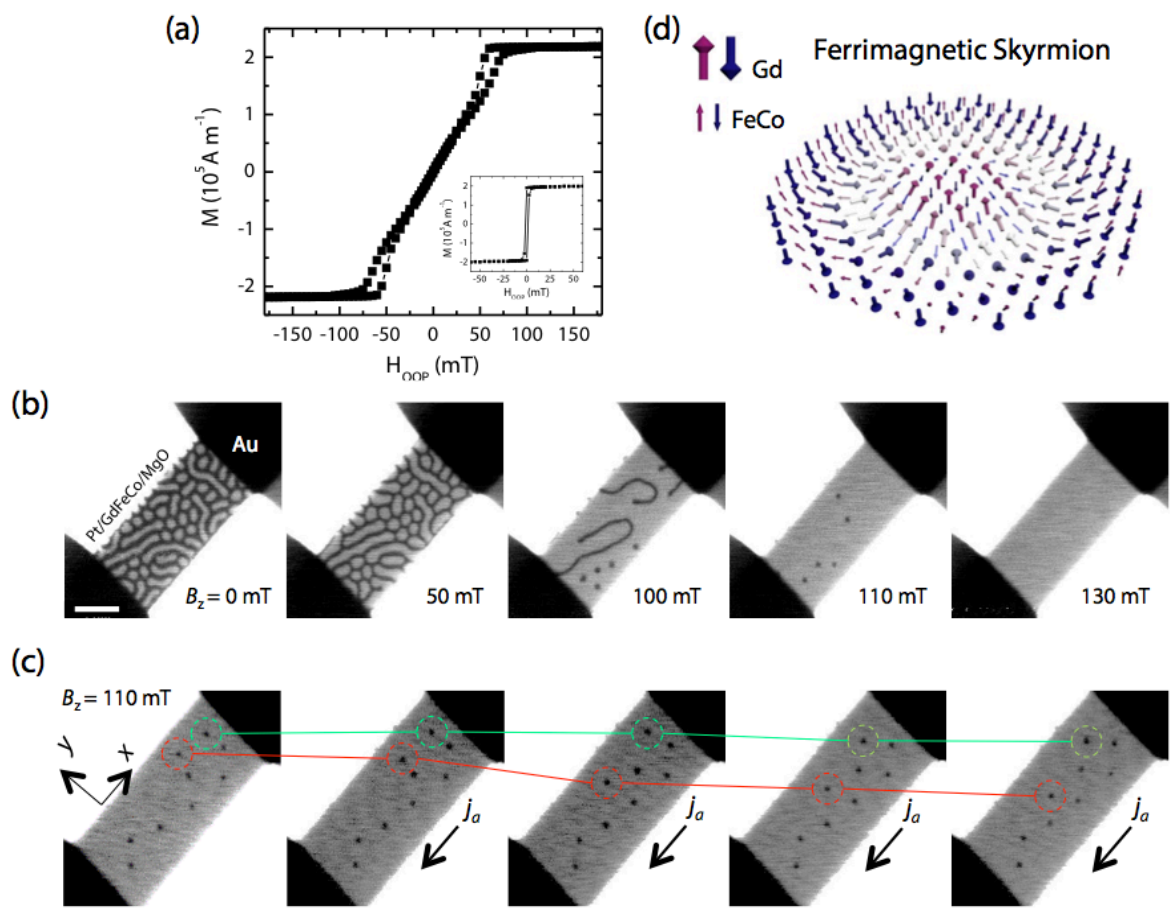

Figure 1

Seonghoon Woo *et al.*

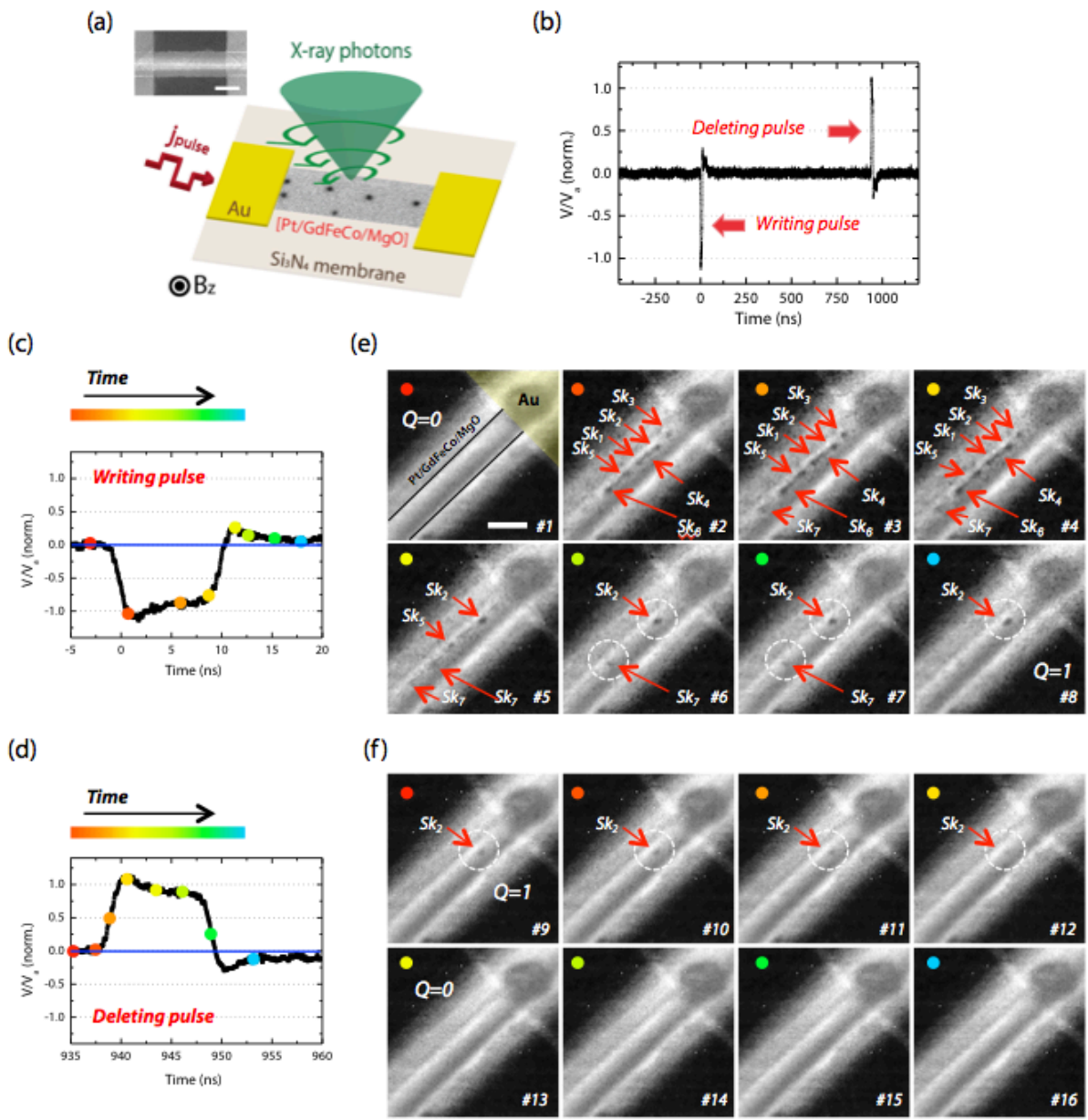

Figure 2

Seonghoon Woo *et al.*

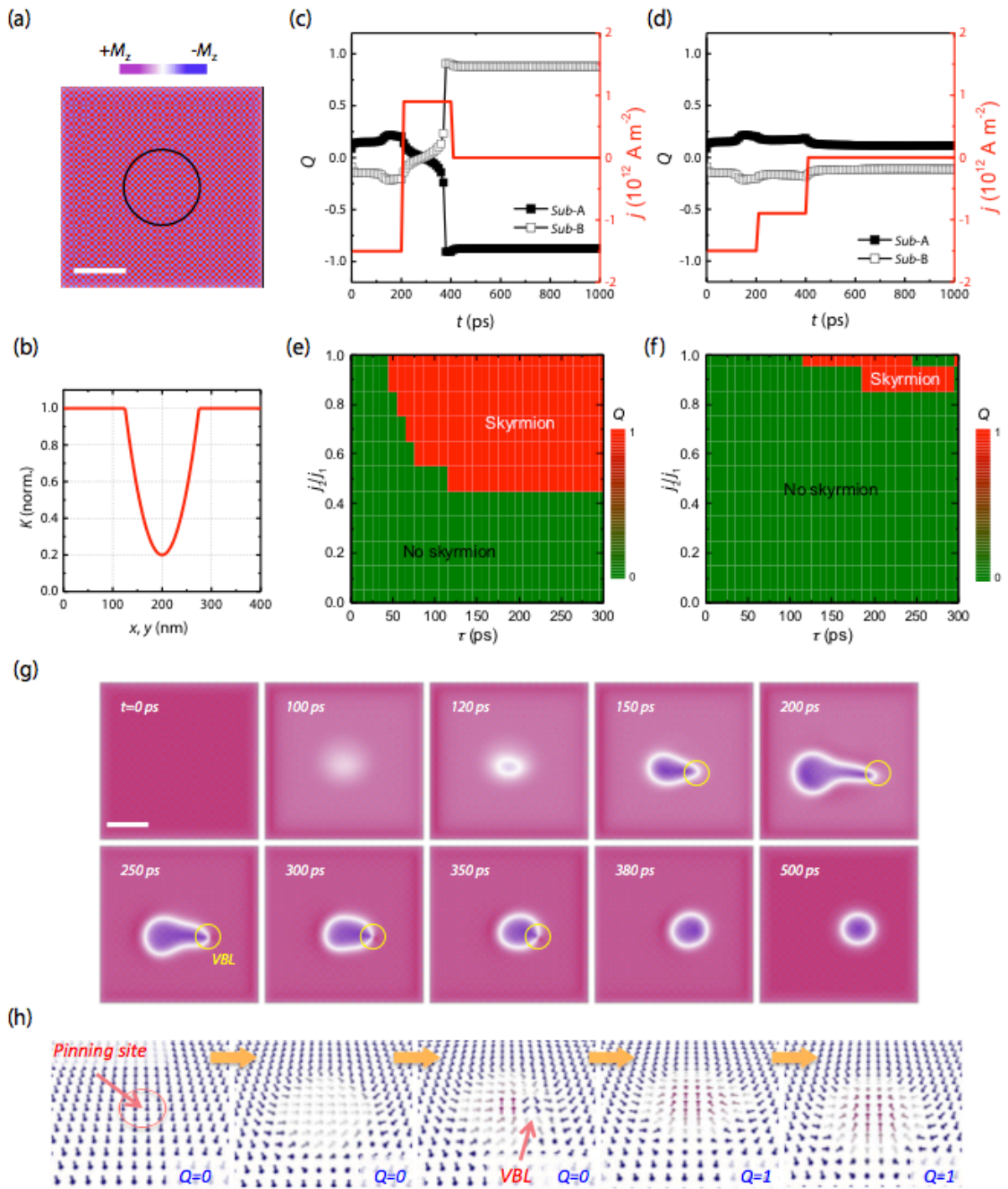

Figure 3

Seonghoon Woo *et al.*

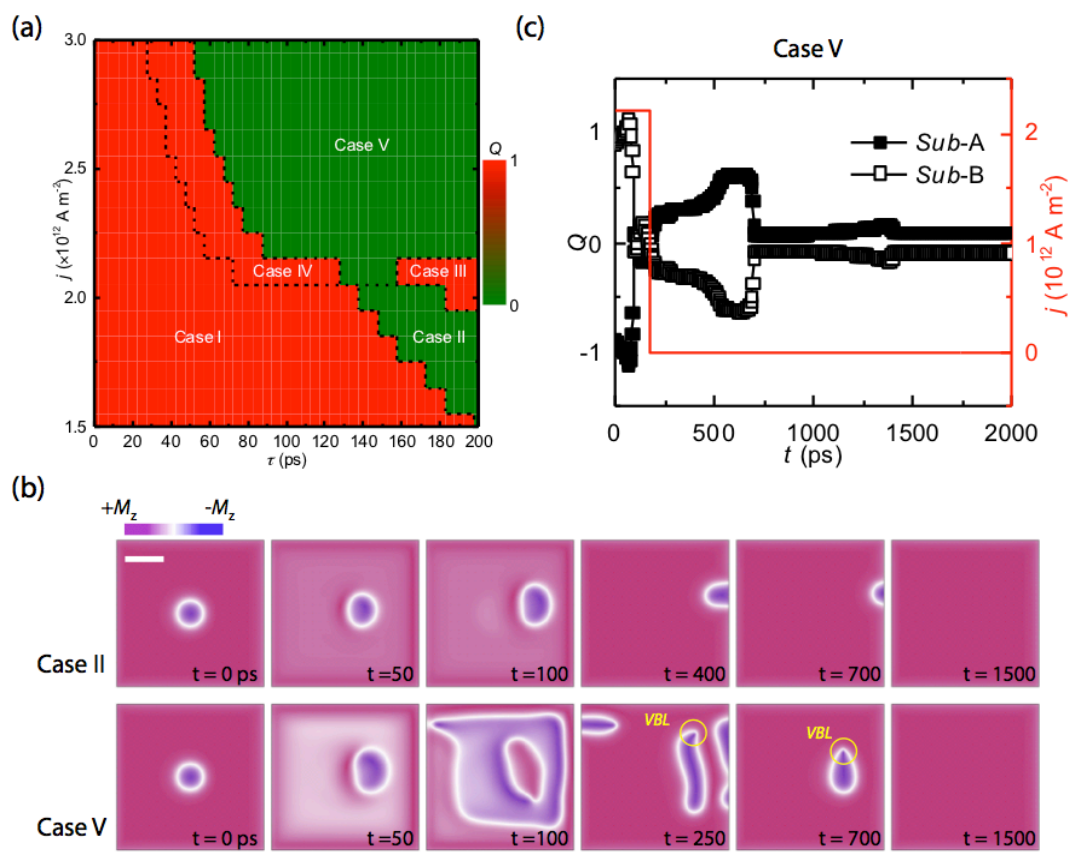

Figure 4

Seonghoon Woo *et al.*